\RequirePackage{ifpdf} 
\ifpdf
\documentclass[pdftex,12pt,a4paper]{article}  
\else
\documentclass[12pt,a4paper]{article} 
\fi

\parskip=\medskipamount

\usepackage[mathscr]{euscript} 	
\usepackage{bbm} 
\usepackage{amsfonts,amsmath,amssymb}
\ifpdf
\usepackage[pdftex]{graphicx}	
\usepackage[usenames,pdftex]{color}		
\usepackage[pdftex,bookmarks]{hyperref}
\hypersetup{bookmarksopen,bookmarksnumbered,
		pdfnewwindow=true,
		pdfauthor = {Simon J. Tyler},
		pdftitle = {Comments on the complex linear Goldstino superfield},
		pdfsubject = {},
		pdfkeywords = {Supersymmetry, Goldstino},
		pdfcreator = {LaTeX with hyperref package},
		pdfproducer = {pdflatex}
}
\else
\usepackage[dvips]{graphicx}
\usepackage[usenames,dvips]{color}
\usepackage[dvips,bookmarks,colorlinks=false]{hyperref}
\fi
\def\a{\alpha}

\def\G{\Gamma}

\def\k{\kappa}

\def\q{\theta}

\def\S{\Sigma} 

\def\X{\Xi}
\newcommand{\eps}{\varepsilon}

\newcommand{\da}{{\dot{\alpha}}}
\newcommand{\db}{{\dot{\beta}}}

\def\ds1{\ensuremath{\mathbbm{1}}}
\newcommand{\rmd}{{\rm d}}

\newcommand{\rmi}{{\rm i}}

\newcommand{\cN}{{\cal N}}

\newcommand{\pd}{\partial}

\def\intx{\int\!\!{\rmd}^4x\,}

\newcommand{\be}{\begin{equation}}
\newcommand{\ee}{\end{equation}}
\newcommand{\bea}{\begin{eqnarray}}
\newcommand{\eea}{\end{eqnarray}}

\newcommand{\Db}{{\bar{D}}}

\begin{document}                        
\begin{titlepage}

\begin{flushright} July, 2015 \end{flushright}

\vspace{5mm}

\begin{center}
{\large \bf  Comments on the complex linear Goldstino superfield}
\end{center}
\begin{center}
{\large  
{Sergei M.\ Kuzenko}
and 
{Simon J.\ Tyler}

\vspace{5mm}

\footnotesize{
{\it School of Physics M013, The University of Western Australia\\
35 Stirling Highway, Crawley W.A. 6009, Australia}}  

\vspace{2mm}}
\end{center}

\vspace{5mm}

\pdfbookmark[1]{Abstract}{abstract_bookmark}
\begin{abstract}
\baselineskip=14pt
It is shown that the complex linear Goldstino superfield 
recently proposed in  arXiv:1507.01885 is obtained from the one originally constructed in 
arXiv:1102.3042 by a field redefinition. 
\end{abstract}
\vfill

\end{titlepage}


Four years ago, we constructed 
the Goldstino model  \cite{KT} described by 
a {\it modified} complex linear superfield $\S$,
\begin{align} \label{mCL:constraint}
	-\frac14 \Db^2 \S = f\,, \qquad f = {\rm const}\,.
\end{align}
Here $f$ is a parameter of mass dimension 2 which, 
without loss of generality, can be chosen to be real.
To describe the Goldstino dynamics, $\S$ was   subject to 
the following nonlinear constraints:
\begin{align}
	\S^2 &= 0~, \label{1st constraint} \\
	-\frac{1}{4} \S\Db^2D_\a\S &= f D_\a\S~. \label{2nd constraint}
\end{align}
The constraint \eqref{1st constraint} tells us that $\S$ is nilpotent.
The form of the Goldstino action coincides with the free action for the complex linear superfield,
\bea
	S[\S,\bar\S] 
	= - \intx\!\rmd^2\q\rmd^2\bar\q\, \S\bar\S  ~.
	\label{Gaction}
\eea

It was also shown in \cite{KT} that all known Goldstino superfields can be obtained as composite constructed from spinor covariant derivatives of
$\S$ and its conjugate $\bar \S$.\footnote{%
This property and the universality  \cite{VA,Ivanov,Uematsu:1981rj}
of the Goldstino \cite{VA} implies that any model for supersymmetry breaking 
can be described in terms of $\S$ and its conjugate.} 
Such constructions make use of
the spinor superfields $\X_\a$
and its conjugate ${\bar \X}_\da$ defined by 
\begin{align} \label{SW_as_Derivative}
\X_\a = \frac{1}{\sqrt2}D_\a \bar \S\,, \qquad
	\bar\X_\da = \frac{1}{\sqrt2}\Db_\da\S\ .
\end{align}
Making use of the constraints \eqref{mCL:constraint}, \eqref{1st constraint} 
and \eqref{2nd constraint}, we can readily uncover 
those constraints which are obeyed by the above spinor superfields. 
They have the form (with  $2\k^2=f^{-2}$)
\begin{subequations}\label{SW_Constraint}
\begin{align}
\label{SW_Constraint1}
	{\bar D}_\da {\bar \X}_\db 
		&= \k^{-1} \eps_{\da \db}\,, \\
\label{SW_Constraint2}
	D_\a {\bar \X}_\da
		&= 2 \rmi \k {\bar \X}^\db \pd_{\a \db} {\bar \X}_\da  
\end{align}
\end{subequations}
and are exactly the constraints given in 
\cite{SamuelWess1983}, so we recognise $\X_\a$ as the Samuel-Wess superfield.
In particular, for the Goldstino superfield $\S$ 
the solution to the constraints \eqref{mCL:constraint}, \eqref{1st constraint} 
and \eqref{2nd constraint} in terms of $\bar \X_\da$ 
is very simple:
\begin{align} \label{mCL-SW-soln}
	2f\S = \bar{\X}^2 \ ,
\end{align}
and the action \eqref{Gaction} takes the form 
\bea
	S[\S,\bar\S] 
	= - \frac{1}{4f^2}\intx\!\rmd^2\q\rmd^2\bar\q\, \X^2 \bar \X^2  ~.
	\label{Gaction2}
\eea

One option for a composite Goldstino superfield that was not discussed in \cite{KT} 
was recently explored in \cite{FHKvU}. It is defined in terms of the Samuel-Wess superfield as
\begin{align}
\Gamma = -\frac{\k^2}{\sqrt{8}} \bar D_\da ( \bar \X^\da \X^2)~.
\end{align}
Making use of \eqref{SW_Constraint} gives
\begin{align}
\G = \frac{\k}{\sqrt{2}} \X^2 \Big\{ 1-\k^2 {\rm i}
\bar \X_\da \partial^{\da \a}\X_\a \Big\}~.
\label{10}
\end{align}
By construction, $\G$ is   an {\it ordinary} complex linear superfield,
\begin{align}
\bar D^2 \G =0\,,
\end{align}
and it follows from \eqref{Gaction2} and \eqref{10} that it has an identical action to $\S$,
\begin{align}
S[\S,\bar\S] = S[\G , \bar \G]~,
\end{align}
where $S[\G , \bar \G]$ is obtained from \eqref{Gaction} 
by replacing $\S \to \G$.
As a consequence of \eqref{10}, $\G$ it is nilpotent,
\begin{align}
\G^2 =0~.
\end{align}
Equations \eqref{SW_Constraint1} and \eqref{SW_Constraint2}
imply that $\G$ satisfies a nonlinear constraint  
that was not discussed  in \cite{FHKvU}
\begin{align}
-\frac14\G\bar D^2\bar\G = f\G ~.
\label{CL-2nd-constraint}
\end{align}
This is structurally identical to the second constraint of Rocek \cite{Rocek},
which is natural since $\G$ is on-shell equivalent to Rocek's constrained chiral superfield \cite{FHKvU}.
As this constraint mixes $\G$ with its conjugate, it leads to complicated solutions for its components.
This is in contradistinction to the constraint \eqref{2nd constraint} that only depends on $\S$.
The solutions to the constraint \eqref{CL-2nd-constraint} and all of the calculations in this paragraph 
can be found in the ancilliary Mathematica notebook attached to \cite{KT0}.

It is interesting to note that in the \emph{modified} complex linear superfield of \cite{KT} 
the constraints mean that the Goldstino arises from the normally physical fermion 
$G_\a \propto \left.D_\a \bar\S\right|$. 
In the complex linear superfield $\G$, the constraints mean that the Goldstino arises 
from a normally auxiliary spinor $G_\a \propto \left.D_\a \G\right|$. 
This was discussed in \cite{FHKvU} and is now easily seen from \eqref{10} 
when rewritten using \eqref{SW_as_Derivative} as 
\begin{align}
\G = \bar\S - \frac{\k}{\sqrt8}(\Db_\da\S)(\Db^\da\bar\S) 	~ .
\end{align}

In conclusion, we would like to point out that the $\S$-realisation \cite{KT}
and the $\G$-realisation \cite{FHKvU} for complex linear
Goldstino superfield are to some extent complementary. 
The former is characterised by the holomorphic-like  constraint  \eqref{2nd constraint} 
that is very useful for applications. The latter has been deduced from
models allowing spontaneous supersymmetry breaking.  
It is worth mentioning that the constraint \eqref{mCL:constraint}
is the only way to describe $\cN=1$ anti-de Sitter supergravity
using a non-minimal scalar multiplet \cite{BK}.


\begin{footnotesize}
\providecommand{\href}[2]{#2}
\begingroup\raggedright
\endgroup
\end{footnotesize}

\end{document}